\documentclass[usenatbib,usegraphicx,twocolumn]{mn2e}
\usepackage[english]{babel} 
\usepackage{color}

\def\th{\vec{\theta}}
\def\u{\vec{U}}

\def\mnras{MNRAS}

\def\aap{A \& A}
\def\apj{Ap.J}

\def\apjl{Ap.JL}
\def\u{{\bf U}} 

\def\V2{V_2}
\def\V2ij{V_{2ij}}
\def\S{{\mathcal S}}
\def\V{\mathcal{V}}
\def\N{{\mathcal N}}

\def\la{\left\langle}

\def\lsim{~\rlap{$<$}{\lower 1.0ex\hbox{$\sim$}}}

\def\gsim{~\rlap{$>$}{\lower 1.0ex\hbox{$\sim$}}}

\usepackage{graphics}
\usepackage{color}
\usepackage{amsmath}
\usepackage{psfrag}
\usepackage{subfig}
\usepackage{graphicx}

\newsavebox{\measurebox}
\begin{document}
\date {} 

\title[Tapering the sky response for $C_{\ell}$ estimation] {Tapering the  sky response for angular power spectrum estimation 
from low-frequency radio-interferometric data}
\author[S. Choudhuri et al.]{Samir Choudhuri$^{1}$\thanks{Email:samir11@phy.iitkgp.ernet.in}, Somnath Bharadwaj$^{1}$\thanks{Email:somnath@phy.iitkgp.ernet.in}, Nirupam Roy$^{1}$, Abhik Ghosh$^{2}$ and
\newauthor Sk. Saiyad Ali$^{3}$\\
  $^{1}$ Department of Physics,  \& Centre for Theoretical Studies, IIT Kharagpur,  Kharagpur 721 302, India\\
$^{2}$ Kapteyn Astronomical Institute, PO Box 800, 9700 AV Groningen, The Netherlands\\
$^{3}$ Department of Physics,Jadavpur University, Kolkata 700032, India}

\maketitle
   
\begin{abstract}
It is important to correctly  subtract point sources from  radio-interferometric 
data  in order to measure the power spectrum of  diffuse  radiation like the 
Galactic synchrotron or the Epoch of Reionization 21-cm signal. It is computationally very expensive and challenging to image a very large area and accurately subtract all the point sources from the image. The problem is 
particularly severe at the sidelobes and the outer parts of the main lobe where 
the antenna response is highly frequency dependent and the calibration also differs 
from that of the phase center. Here we show that it is possible to overcome this problem 
by tapering  the sky response. 
 Using simulated  $150 \, {\rm MHz}$ observations, 
we demonstrate that it is possible to suppress the contribution due to point sources from the outer parts
by using the Tapered Gridded Estimator to measure the angular power spectrum $C_{\ell}$
of the sky signal. We also show from the simulation that this method can self-consistently compute the noise bias and accurately subtract it to provide an unbiased estimation of $C_{\ell}$.

\end{abstract} 


\section{Introduction}
 Foreground removal for detecting the Epoch of
Reionization (EoR) 21-cm signal is a topic of intense current research 
\citep{jelic08,bowman,paciga11,chapman12,liu2,mao12,paciga13}.
Foreground avoidance
\citep{datta10,parsons12,trott1,vedantham12,pober13,thyag13,parsons14,dillon14,pober14,liu14a,liu14b,zali15} 
is an alternate strategy based on the proposal  that the 
foreground contamination is restricted to a wedge  in
$(k_{\perp},k_{\parallel})$ space, and the signal can be estimated from the
uncontaminated modes outside the wedge. 
Point sources dominate the $150 \, {\rm MHz}$ sky  at the
angular scales $\le  4^{\circ}$ \citep{ali08} which are relevant 
for  telescopes like the Giant Metrewave Radio Telescope
\citep[GMRT;][]{swarup91}, Low-Frequency Array
\citep[LOFAR;][]{van13} and the  upcoming Square Kilometre Array\footnote{https://www.skatelescope.org} (SKA). 
It is difficult to model and subtract the point sources at 
the periphery of the telescope's field of view. The difficulties include the
fact that the antenna response  is highly frequency dependent near the nulls
of the primary beam,  and  the calibration differs from that of the
phase center due to ionospheric fluctuations.
Point source subtraction is also
important for measuring the angular power spectrum of the diffuse Galactic
synchrotron radiation \citep{bernardi09,ghosh150,iacobelli13}   which, apart 
from being an important foreground component for the EoR 21-cm signal,
is interesting in its own right. 

Most of the foreground subtraction techniques use the property of  smoothness
 along frequency for the various foreground components. \citet{ghosh1,ghosh2}
 found  that  residual point sources 
 located away from the phase center introduce oscillations  along frequency direction. 
The oscillation  are more rapid if the distance of the source from the phase 
center increases, and also with increasing baseline. Equivalently, the dominant
contribution to the width of the foreground wedge arises from the sources
located at the periphery of the field of view \citep{thyag13}.  
Using GMRT \citet{ghosh2,ghosh150} have shown that these oscillations can be reduced by tapering the sky response. In a recent paper \citet{pober16} showed that correctly modelling and subtracting the sidelobe foreground contamination is important for detecting the redshifted 21-cm signal.

In a recent paper \citet{samir14} have introduced the Tapered Gridded Estimator 
(TGE) for estimating the angular power spectrum $C_{\ell}$ directly from
radio-interferometric visibility data.  In this paper we use simulated
$150 \, {\rm MHz}$ GMRT data which incorporates point sources and the diffuse
Galactic synchrotron radiation to demonstrate that it is possible to suppress
the contribution from residual point sources in the sidelobes and the outer parts of the primary 
beam in estimating $C_{\ell}$ using the TGE. 

Noise bias  is an important issue for any estimator. For example, the image based estimator
\citep{seljak97} for $C_{\ell}$ and the visibility based estimator \citep{liu2}
for $P(k_{\perp},k_{\parallel})$ rely on modelling the noise properties of the
data and subtracting out the expected noise bias. However, the actual noise in
the observations could have baseline, frequency and time dependent  variations
 which  are  very difficult to model and there is the  risk of residual noise
 bias being mistaken as the signal. \citet{paciga11} have
avoided the noise bias by cross-correlating observations made on different
days. Another visibility based estimator \citep{begum06,prasun07}
individually correlates pairs of visibilities avoiding
the self correlation that is responsible for the noise bias. This, however, is
computationally  very expensive when the data volume is large. In this paper, we have
demonstrated that TGE, by
construction, estimates the actual noise bias internally from the data and
exactly subtracts this out to give an unbiased estimate of $C_{\ell}$. The entire discussion here is in the context of estimating  $C_{\ell}$ for the
diffuse Galactic synchrotron radiation. As mentioned earlier,  the same issues
are also relevant for measuring the EoR 21-cm power spectrum not considered here. 

In Section~\ref{sec:imgproblem} we discuss the conventional problem in standard imaging techniques. Simulation and data analysis processes are briefly discussed in Section~\ref{simu}.  Section~\ref{sec:est} discusses the estimator (TGE) that we used to suppress the outer region of the primary beam and the results are presented in Section~\ref{result}. Finally, we present summary and conclusion in Section~\ref{sum}.

\section{Problems in conventional Imaging}
\label{sec:imgproblem}
The contribution to the signal in radio frequency observations from the outer region of the primary beam and from the sidelobes is generally very small as compared to the inner region of the primary beam. In particular, the expected 21-cm signal, which itself is very faint, contributes mainly from the central part of the primary beam, and attenuated to a great extent in the outer region. Only the bright point sources from the outer region, if not accurately removed, may have significant impact on the statistical estimation of the diffuse signal. Thus, it is necessary to remove the effect of point sources from the outer region before estimating the residual power spectrum. However,  we will not be benefitted in terms of signal by including highly attenuated diffuse emission from the outer region.

Imaging a large enough region to model and subtract all the point sources before dealing with the diffuse emission may seems to be a direct solution of the above problem. But, in reality there are many issues which make this approach impractical. First of all, the field of view at low radio frequencies is large, and making larger images is computationally more expensive. In addition to that, non-coplaner nature of the baselines prevents us from making wide-field image without considering the effect of the ``w-term''. There are algorithms e.g. faceting \citep{cornwell92}, w-projection \citep{cornwell08}, WB-A projection \citep{bhat13} etc. to tackle this problem partly for radio interferometric observations. However, these algorithms still require significant computation to make an image of such a large region of the sky. Secondly, the number of bright point sources is quite large at low frequency. While imaging a very large region, selecting CLEANing region around each source is a tedious job. On the other hand, CLEANing without selecting regions removes a non-negligible part of the diffuse signal of our interest \citep[see][for details]{samir15}. 

The next challenge is to accurately characterize the time and frequency dependence of the wide-field primary beam for effective point source subtraction from the periphery of the telescope's field of view \citep[e.g.][]{neben}. Both the frequency dependence and the deviation from circular symmetry are more prominent at the outer part of the primary beam. These, along with the rotation of primary beam on the sky, cause a strong time and frequency variation of the primary beam for point sources in the outer region. They create problem in accurately model the point sources that we want to subtract from the data. In fact, some of the variations are intractable in nature and  it is extremely difficult, if not impossible, to make accurate modelling and subtraction of the point sources from the outer part of  the primary beam.

Though we have not considered instrumental gains and ionospheric effects in this study, in real life any directional dependence of these quantities will also severely limit our ability to subtract point sources accurately from a large region. One can overcome this difficulty to some extent by going into complicated and messy procedure of direction dependent calibration (e.g. peeling) \citep{bhat08,intema09,kazemi11}. Again, (a) it is computationally more expensive, (b) part of the variation may be intractable, and (c) there is hardly any gain in terms of recovering the diffuse signal which is too weak in outer region. 

The future generation low frequency telescopes (e.g. SKA) that will presumably be used to carry out redshifted diffuse H~{\sc i} observation, will have larger field of view, large bandwidth, longer baseline and higher sensitivity. Hence the above issues will be even more relevant. Moreover, the expected huge data volume from observations with those telescopes will make it more challenging to address these problems by imaging a larger region for subtracting the point sources.
The following two sections outline a technique to overcome these problems by subtracting point sources only from the central region and using the TGE to recover the power spectrum of the diffuse emission in a more efficient way.

\section{Simulation and Data Analysis}
\label{simu}
The details of the simulation and data analysis,  including point source
subtraction, are presented in a companion paper \citep{samir15} and we only
present a brief discussion here. Our  model of the $150 \, {\rm MHz}$ sky has
two components, the first being the diffuse Galactic synchrotron radiation
which is the signal that we want to detect. We use the measured 
angular power spectrum  \citep{ghosh150} 
\begin{equation}
C^M_{\ell}(\nu)=A_{\rm 150}\times\left(\frac{1000}{\ell} \right)^{\beta}\times\left(\frac{\nu}{150{\rm MHz}}\right)^{-2\alpha}   \,.
\label{eq:cl150}
\end{equation}
as the input model to generate  the brightness temperature fluctuations  on
the sky. Here  $\nu$ is the frequency in ${\rm MHz}$, $A_{\rm 150}=513 \, {\rm
  mK}^2$, $\beta=2.34 $  \citep{ghosh150} and $\alpha=2.8$ \citep{platinia98}. 
  The simulation covers a $\sim 8^{\circ} \times \sim 8^{\circ}$
region of the sky and a $16 \, {\rm MHz}$ bandwidth, centered at $150 \, {\rm MHz}$,  over $128$ spectral channels. The diffuse signal was simulated on a grid of
resolution $\sim 0.5^{'}$. 

The Poisson fluctuation of the extragalactic point sources dominates the $150
\, {\rm MHz}$ sky at the angular scales of our interest \citep{ali08}, and it is necessary
to subtract these or suppress their contribution in order to detect any 
  diffuse component like the Galactic synchrotron radiation which we consider here or
  the redshifted 21-cm cosmological signal which is much fainter and is not
  considered here. 
We use the $150\, {\rm MHz}$ differential source count measured  using GMRT 
\citep{ghosh150}
\begin{equation}
\frac{dN}{dS} = \frac{10^{3.75}}{Jy \cdot Sr}\cdot\,\left(\frac{S}{1
  Jy}\right)^{-1.6} \,.
\label{eq:b1}
\end{equation}
 to generate point sources in the flux range  $9 {\rm mJy}$ to 
$1 {\rm Jy}$ whose angular positions are randomly distributed 
within the $3.1^{\circ} \times 3.1^{\circ}$ Full Width Half Maxima (hereafter FWHM)  of the primary beam. 
The antenna response falls off beyond the FWHM, and we only include the
bright sources ($S \ge   100 {\rm mJy}$) outside the FWHM. We have $353$
and $343$ sources in the inner and outer regions respectively, and the sources
were assigned a randomly chosen spectral index $\alpha$ ($S_{\nu} \propto
\nu^{-\alpha}$) in the range $0.7$ to $0.8$.

\begin{figure}
\begin{center}
\psfrag{theta}[c][c][1.][0]{$\theta$ $[{\rm Degrees}$]}
\psfrag{Pbeam}[c][c][1.][0]{${\mathcal A}(\th)$}
\psfrag{Bessel}[r][r][1.][0]{Primary Beam}
\psfrag{taper1}[r][r][1.][0]{${\cal A_W}(\th)$,f=2.0}
\psfrag{taper2}[r][r][1.][0]{${\cal A_W}(\th)$,f=0.8}
\psfrag{taper3}[r][r][1.][0]{${\cal A_W}(\th)$,f=0.6}
\includegraphics[width=80mm,angle=0]{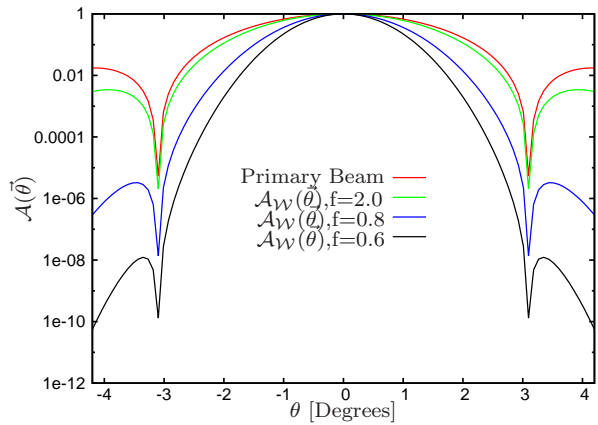}
\caption{The GMRT $150 \, {\rm MHz}$ primary beam  ${\cal A}(\th)$  
which has been modelled as the  square of a Bessel function.  The effective primary beam 
${\cal A_W}(\th)$, obtained after tapering the sky response for the 
different values of $f$ is also shown in the figure.} 
\label{fig:taper}
\end{center}
\end{figure}

We consider the mock GMRT observations 
targeted on a arbitrarily selected  field located at RA=$10{\rm h} \, 46{\rm
  m} \, 00{\rm s}$ and DEC=$59^{\circ} \, 00^{'} \, 59^{''}$. The GMRT has $30$
 antennas which for 
a total  $8 \, {\rm hr}$ of observation with  $16 {\rm s}$ integration time 
results in $783,000$ baselines $\u_i$ with $128$ visibilities $\V(\u_i,\nu)$
 (one per frequency channel) for  each baseline. The resolution of GMRT at $150 \, {\rm MHz}$ is $20^{''}$. 
The diffuse signal (eq. \ref {eq:cl150}) falls off with  increasing $U=\mid \u \mid$ 
($\ell = 2 \pi U$), and we  include this contribution for only the small baselines $U \le 3,000$
for which the visibility contribution is calculated using a 2 dimensional Fourier transform. 
We note that the $w$ term does not significantly affect the diffuse signal \citep{samir14}, 
however this is very important for correctly imaging and subtracting the point sources. 
We have included the point source contribution  for all the baselines in the simulation, and 
the visibilities are  calculated by individually summing over each point source  and including 
the $w$ term. We have modelled the GMRT primary beam pattern ${\cal A}(\th,\nu)$ 
 with the  square of a Bessel function (Figure \ref{fig:taper}) corresponding to the telescope's $45 \, {\rm m}$
diameter circular aperture. The simulated sky is multiplied with ${\cal A}(\th,\nu)$  
before calculating  the visibilities. Finally, we add the system noise contribution which is modelled 
a Gaussian random variable 
with standard deviation  $\sigma_n=1.03{\rm Jy}$ for  the real and imaginary parts of each visibility. 
We note that the GMRT has two polarizations which have identical sky signals but independent noise. 
 
\begin{figure}
\begin{center}
\psfrag{a}[c][c][1.5][0]{{$\hspace{5mm}$ }}
\psfrag{b}[c][c][1.5][0]{{$\hspace{5mm}$ }}
\psfrag{x}[c][c][0.6][0]{$\theta_x$ $[{\rm Degrees}$]}
\psfrag{y}[c][c][0.6][0]{$\theta_y$ $[{\rm Degrees}$]}
\includegraphics[width=75mm]{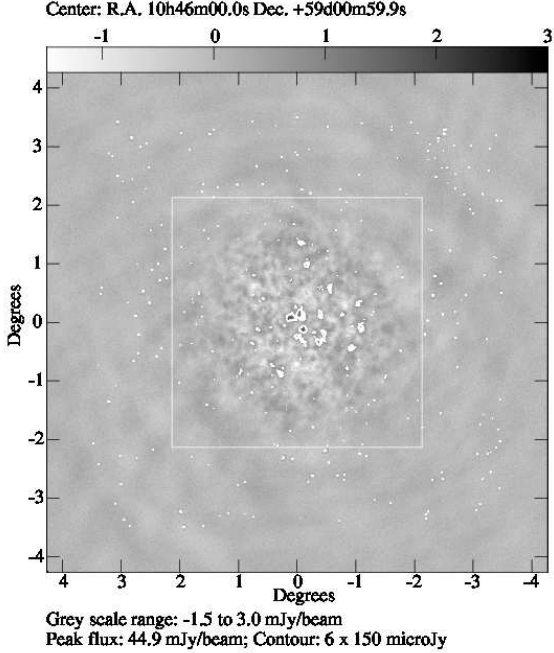}
\caption{``Dirty'' image of the entire simulation region made with
  the residual visibility data after point source subtraction. Point sources
  were subtracted from a central region (shown with a box, $4.2^{\circ}\times4.2^{\circ}$) whose extent is
  $\sim 1.3$  times the FWHM of the primary beam. The features visible inside
  the box all correspond to the diffuse radiation.  Residual point sources are
  visible outside the box, however the diffuse radiation is not visible  in
  this region.}
\label{fig:img}
\end{center}
\end{figure}

We have used the  Common Astronomy Software Applications (CASA)\footnote{http://casa.nrao.edu/} 
package to image and analyze our simulated data. The standard tasks  CLEAN and UVSUB were used to model 
and subtract out the point sources from a $4.2^{\circ}\times4.2^{\circ}$ region which covers an extent 
that is   approximately $1.5$  times the FWHM of the primary beam. We have tried different CLEAN strategies
for which the details are presented in our companion paper \citep{samir15}, and for this work we adopt 
the most optimum parameter values which correspond to Run(e) of the companion paper.  
Figure~\ref{fig:img} shows the  ``dirty'' image of the entire simulation  region 
made from the residual visibility data after point source subtraction. The
central square box ($4.2^{\circ}\times4.2^{\circ}$) shows the region from which  the point sources have been  
subtracted. The features visible in this region correspond to the Galactic synchrotron radiation. 
It is difficult to model and subtract point sources from the periphery where the antenna response is 
highly frequency dependent. It also needs creating and cleaning a huge image that is computationally more expensive. Further, in real observations, any direction dependent gain away from the phase center will make it even more difficult. We have not attempted to subtract the point sources
from the  region outside the central box and  the residual point sources are
visible in this region of the  image.

Figure~\ref{fig:comtaper}  shows the angular power spectrum $C_{\ell}$ before and after point source subtraction;
the input  model for the diffuse radiation is also shown for comparison. 
Before subtraction, the point sources dominate $C_{\ell}$ at all angular multipoles $\ell$.  
After subtraction, we are able to recover the diffuse component at low angular multipoles
 $\ell\le3\times10^{3}$. However, the residual point sources still dominate at the  large 
$\ell$ values. The goal is to suppress the contribution from the residual point sources 
located at the periphery of the beam so that we can recover the input model over the entire $\ell$ 
range.  We show that it is possible to achieve this with the Tapered Gridded
Estimator discussed in the next section.

\section{The Tapered Gridded Estimator}
\label{sec:est}
The observed  visibilities are  a sum of  two independent parts  namely the  sky signal and the system noise 
\begin{equation}
\V(\u, \nu)=\S(\u, \nu)+\N(\u,\nu) \,.
\label{eq:c1}
\end{equation}
The signal $\S(\u, \nu)$ and the noise $\N(\u,\nu)$ are considered to be
  independent random variables, further the noise in the different
  visibilities are uncorrelated.
 The signal  contribution $\S(\u,\nu)$ records the Fourier transform of the product of 
$\delta I(\th,\,\nu)$, the fluctuation in specific  intensity of the sky signal, and 
the telescope's primary beam pattern ${\mathcal A}(\theta, \nu)$ shown in Figure \ref{fig:taper}. 
As mentioned earlier, it is difficult to model  and subtract point sources from the outer
region of the primary beam and the sidelobes. The residual point sources in the periphery 
of the telescope's field of view pose a problem for estimating the power spectrum of 
the diffuse radiation.  
In this section we discuss the Tapered  Gridded Estimator (TGE) 
which is  a technique for estimating the angular power spectrum from the visibility data.  
This technique suppresses the
contribution from  the sidelobes and the outer part of the primary beam by tapering the sky response. 
\citet{samir14} presents a detailed discussion of this estimator, and we only present a brief outline
 here. 

We  taper the sky response by multiplying the  field of view with a frequency independent 
Gaussian window function 
${\cal W}(\theta)=e^{-\theta^2/\theta^2_w}$. Here we parametrize $\theta_w=f \theta_0$ where 
$\theta_0=0.6 \times \theta_{FWHM}$ and $\theta_{FWHM}$ is the FWHM of the 
telescope's primary beam at the central frequency, and preferably $f \le 1$ so that ${\cal W}(\theta)$
  cuts off the sky response well before the first null of the  primary beam. 
We implement the tapering by convolving the measured visibilities with $\tilde{w}(\u)$ 
the Fourier transform of ${\cal W}(\theta)$. The convolved visibilities are evaluated 
on a grid in $uv$ space using 
\begin{equation}
\V_{cg} = \sum_{i}\tilde{w}(\u_g-\u_i) \, \V_i
\label{eq:c5}
\end{equation}
where $\u_g$ refers to the different grid points and $\V_i$ 
refers to the measured visibilities at baseline $\u_i$. The gridding significantly reduces
the data volume and the computation time required to estimate the power spectrum \citep{samir14}. 
The convolved visibilities are calculated separately  for each frequency channel. 
Then, for the purpose of this work, convolved visibilities for a grid are averaged over all frequencies. 

The  signal component of the convolved visibility is the Fourier transform of the product of a  
 modified   primary beam pattern ${\mathcal A_W}(\th, \nu)={\cal W}(\theta)\,  {\cal A}(\th, \nu)$
and $\delta I(\th, \nu)$
\begin{equation}
\S_c(\u,\nu)=  \int \, d^2 \th  \,
     {\mathcal A_W}(\theta, \nu)\delta I(\th,\,\nu)e^{2\pi i \u.\th} \,.
\label{eq:c3}
\end{equation}
It is clear that the convolved visibilities respond to the signal from a smaller region of the sky 
as compared to the measured visibilities. It may be noted that 
the tapering is effective only if the window function $\tilde{w}(\u_g-\u_i)$
in eq. (\ref{eq:c5})  is well sampled by  the   baseline distribution. 
The results of this paper, presented later, indeed justify this assumption for
the GMRT.

The correlation of the gridded visibilities $\langle \V_{c g}  \V^{*}_{c g} \rangle$ 
gives a direct  estimate of the angular power spectrum $C_{\ell_g}$ through 
\begin{equation}
\langle \V_{c g}  \V^{*}_{c g} \rangle
 = \mid K_{1g} \mid^2V_1 C_{\ell_g} + \sum_i  \mid \tilde{w}(\u_g-\u_i)
 \mid^2 \langle  \mid \N_i \mid^2 \rangle  
\label{eq:c6}
\end{equation}
where the angular multipole $\ell_g$ is related to the baseline $U_g$ as 
$\ell_g=2 \pi U_g$, $K_{1g}=\sum_i  \tilde{w}(\u_g-\u_i)$, 
$V_1= \left( \frac{\partial B}{\partial T}\right)^{2}\left[\int d^2 U{'} \,  
\mid \tilde{a}_W(\u-\u{'}) \mid^2 \right]$,
$\tilde{a}_W$ is the Fourier transform of ${\mathcal A_W}$
and $\left( \frac{\partial B}{\partial T}\right)$ is the conversion factor from 
brightness temperature to specific intensity. We see that the visibility correlation 
also has a term  involving $ \langle  \mid \N_i \mid^2 \rangle$
which is the variance of the noise contribution  present in the 
measured visibilities (eq. \ref{eq:c1}). 
This term, which   is independent of $C_{\ell}$, introduces a positive
definite noise bias.    The visibility correlation (eq. \ref{eq:c6}) provides
an estimate of $C_{\ell}$ except for the  additive noise bias. 
The TGE uses the same visibility data to obtain an  internal estimate of the
noise bias and subtract it from the visibility correlation. 
 We  consider the self-correlation term $B_{cg}=\sum_i \mid \tilde{w}(\u_g-\u_i) \mid^2 \,
\mid  \V_i \mid^2 $ for which 
\begin{equation}
 \langle B_{cg} \rangle
=   \sum_i \mid  \tilde{w}(\u_g-\u_i)\mid^2  (V_0  C_{\ell_i} 
+ \langle  \mid \N_i \mid^2 \rangle ) \,.
\end{equation}
where $V_0= \left( \frac{\partial B}{\partial T}\right)^{2}\left[\int d^2 U{'} \,  
\mid \tilde{a}(\u-\u{'}) \mid^2 \right]$,
$\tilde{a}$ is the Fourier transform of the primary beam pattern ${\mathcal
  A}$.  
 The term  $ \langle B_{cg} \rangle$, by construction, has  exactly the same
 noise bias as the visibility 
 correlation in eq. (\ref{eq:c6}). We use this to define the 
TGE estimator  
\begin{equation}
{\hat E}_g= (\mid K_{1g} \mid^2 V_1)^{-1} [ \V_{c g}  \V^{*}_{c g} - B_{cg}]
\end{equation}
which gives an unbiased estimate of the angular power spectrum at a grid point $g$.
A part of the signal also gets subtracted out with the noise bias. This loss is proportional to 
$N$ (the number of visibility data) whereas the visibility correlation is
proportional to $N^2$, 
and this loss is insignificant when the data size is large \citep{samir14}.   
The $C_{\ell_g}$ values estimated at each grid point  are  binned  in
logarithmic intervals of $\ell$,  
and we consider the bin-averaged $C_{\ell}$ as a function of the bin-averaged
angular multipole $\ell$.  

Tapering reduces the sky coverage which, in addition to suppressing the
point sources in  the periphery of the main lobe and the sidelobes, also
affects the diffuse signal. The reduced sky coverage causes the cosmic
variance of the estimated  $C_{\ell}$ to increase as $f$ is reduced (Figure
10, \citealt{samir14}). Further, the reduced sky coverage also restricts the 
 $\ell$ range ($\ell_{min} - \ell_{max}$) where it is possible to estimate
$C_{\ell}$,  and the value  of $\ell_{min}$ increases as $f$ is decreased.
\section{Results}
\label{result}
We have applied the Tapered Gridded Estimator (TGE) to the residual visibility data  after subtracting out the point sources. 
As mentioned earlier, the point sources have been identified and subtracted from  a $4.2^{\circ}\times4.2^{\circ}$ region  
(Figure~\ref{fig:img}) which covers an extent that is $\approx 1.3$ times
the FWHM   of the primary beam. However, the  point sources still remain 
at the periphery of the primary beam  and in the part of the sidelobe which has been included in the simulation.   The TGE tapers
the sky response which results in  an  effective primary beam ${\cal A_W}(\th)$  that  is  considerably narrower than the 
actual primary beam of the telescope ${\cal A}(\th)$.  Figure \ref{fig:taper} shows ${\cal A_W}(\th)$ for three different values of $f$ 
($2.0,0.8$ and $0.6$). For $f=2.0$  we see that ${\cal A_W}(\th)$ is not very significantly different from ${\cal A}(\th)$ in the region 
within  the first  null,  the difference however increases in the first sidelobe  and the sidelobe response is suppressed by a factor
 of $10$  at $\mid \th \mid \approx 4^{\circ}$. We see that the effective primary beam  gets narrower as the value of $f$ is reduced. 
The value of ${\cal A_W}(\th)$ is a factor of $\approx 10$ ($100$) lower compared to ${\cal A}(\th)$ for $f=0.8$ ($0.6$)
at  $\mid \th \mid = 2^{\circ}$ which corresponds to  the boundary of the region  within which the point sources have been subtracted.
We see that, for $f=0.8$ ($0.6$), tapering suppresses the first side lobe of ${\cal A_W}(\th)$ by a factor of $\approx 10^{5}$ ($10^{8}$) compared to
${\cal A}(\th)$  at  $\mid \th \mid = 4^{\circ}$. We expect the residual point source contribution to reduce by at least a factor of $10$ and $100$ for $f=0.8$ and $0.6$ respectively. 

\begin{figure}
\begin{center}
\psfrag{cl}[b][t][1.5][0]{$C_{\ell} [mK^2]$}
\psfrag{U}[c][c][1.5][0]{$\ell$}
\psfrag{model}[r][r][1][0]{Model}
\psfrag{allnotaper}[r][r][1][0]{Total}
\psfrag{notaper}[r][r][1][0]{Residual}
\psfrag{tap2}[r][r][1][0]{f=2.}
\psfrag{tap0.8}[r][r][1][0]{f=0.8}
\psfrag{tap0.6}[r][r][1][0]{f=0.6}
\includegraphics[width=80mm,angle=0]{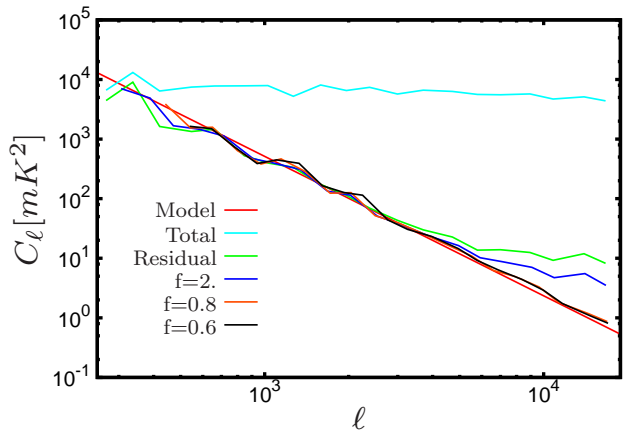}
\caption{Angular power spectrum $C_{\ell}$ of total and residual data. It also shows the estimated $C_{\ell}$ using  the TGE for the different values of $f$ are also shown in the figure. In this figure the curves for $f=0.6$ and $0.8$ overlaps with each other.}
\label{fig:comtaper}
\end{center}
\end{figure}

Figure \ref{fig:comtaper} shows the angular power spectrum ($C_{\ell}$) estimated from the residual visibility 
data using TGE with the $f$ values ($2.0,0.8$ and $0.6$) discussed earlier.  We see that in the absence of tapering 
we are able to recover the  angular power spectrum of the diffuse synchrotron radiation at  the low 
angular multipoles (large angular scales) $\ell < 3\times10^3$. The residual point source 
contribution is nearly independent of $\ell$ and has a value $C_{\ell} \approx 10$ ${\rm mK}^2$ which 
dominates the estimated $C_{\ell}$ at the large angular multipoles (small angular scales)  
$\ell \ge 10^4$. We have a gradual transition from the diffuse synchrotron dominated to a point 
source dominated $C_{\ell}$ in the interval $ 3\times10^3 \le \ell < 10^4$. The point source 
contribution comes down by a factor of  more than  $2$  if we use the TGE with $f=2.0$. 
We are now able to recover the  angular power spectrum of the diffuse synchrotron 
radiation to larger $\ell$ values ($\ell < 5\times10^3$) as compared to the situation 
without tapering.  The point source contribution, however, still dominates at larger $\ell$ 
values. We find that the point source contribution to $C_{\ell}$ is suppressed by more than a 
factor of $10$ if we use TGE with $f=0.8$ or $0.6$.  We are able to recover the 
angular power spectrum of the diffuse synchrotron radiation  over the entire $\ell$ range 
using either value of $f$. The fact that there is no noticeable change in  $C_{\ell}$
if the value of  $f$ is reduced from $0.8$ to $0.6$ indicates that a tapered
sky response with $f=0.8$ is adequate 
to detect the angular power spectrum of 
the diffuse synchrotron radiation  over the entire $\ell$ range of our interest here. 

\begin{figure}
\begin{center}
\psfrag{cl}[b][t][1.5][0]{$C_{\ell} [mK^2]$}
\psfrag{U}[c][c][1.5][0]{$\ell$}
\psfrag{model}[r][r][1][0]{Model}
\psfrag{0.8nonse}[r][r][1][0]{No Noise Bias}
\psfrag{0.8nse}[r][r][1][0]{Noise Bias}
\includegraphics[width=80mm,angle=0]{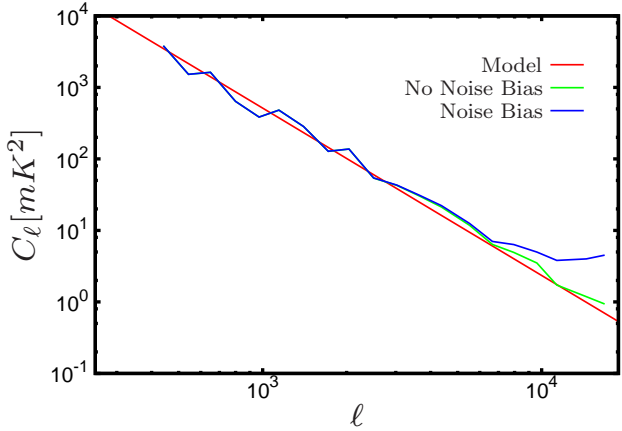}
\caption{Angular power spectrum $C_{\ell}$ estimated using  the TGE with $f=0.8$.
Results with the noise bias being present and with the noise bias subtracted are both shown here. }
\label{fig:comnse}
\end{center}
\end{figure}

The noise bias is an important issue in estimating the angular power spectrum, we illustrate 
this in Figure \ref{fig:comnse}. For this purpose we have used a smaller frequency bandwidth 
of $8 \ {\rm MHz}$ which increases the noise r.m.s. compared to the 
$16 \ {\rm MHz}$ bandwidth used throughout the rest of the paper. Figure \ref{fig:comnse} shows 
$C_{\ell}$ estimated with the TGE with $f=0.8$. We expect to recover the 
angular power spectrum of the diffuse synchrotron radiation  over the entire $\ell$ range  provided the
 noise bias is correctly estimated and subtracted out. 
 Figure \ref{fig:comnse} shows the estimated $C_{\ell}$  in the situation where the noise bias 
is not subtracted. We see that the noise bias makes a nearly constant contribution of 
$C_{\ell} \approx 7.5$ ${\rm mK}^2$ which dominates the estimated $C_{\ell}$ at large $\ell$. 
It is necessary to subtract the noise bias in order  to recover the $C_{\ell}$
of the diffuse  radiation at large $\ell$.   Figure \ref{fig:comnse}  demonstrates 
that the  TGE correctly subtracts out  the noise bias so that we are  able to recover the 
$C_{\ell}$  of the diffuse  radiation  over the entire $\ell$ range.

\section{Summary and Conclusion}
\label{sum}
It is difficult to model and subtract point sources located at the periphery of the 
telescope's field of view. These residual point sources pose a problem for estimating 
the power spectrum of the diffuse background radiation if all visible point sources are removed with high level of accuracy from inside the main lobe of the primary beam. For example, \citet{pober16} have recently shown the effect of the residual point sources outside the main lobe on estimating the power spectrum for MWA observation. This issue is discussed here in 
the context of measuring the angular power spectrum of the diffuse Galactic synchrotron radiation 
using simulated $150 \, {\rm MHz}$ GMRT observations. However, the same  issue is also very 
important for  detecting the EoR 21-cm power spectrum which is a much 
fainter  diffuse signal that is not considered here. 

It is possible to suppress the contribution from the residual
point sources located at the periphery of the telescope's field of view through  a 
frequency independent window function  which  restricts or tapers the sky response. 
The Tapered Gridded Estimator(TGE)  achieves this tapering by convolving the measured visibilities 
with the Fourier transform of the window function. This estimator for the angular 
power spectrum has the added advantage that it internally estimates the noise bias
from the measured visibilities and accurately subtracts this out to provide an unbiased
estimate of  $C_{\ell}$.  In this paper we demonstrate, using simulated data, that the 
 TGE very effectively suppresses the contribution of the residual
point sources located at the periphery of the telescope's field of view. We also
 demonstrates that the TGE correctly estimates the noise bias from the input 
visibilities  and subtracts this out to give an unbiased estimate of $C_{\ell}$. 

The issues considered here are particularly important in the context of measuring the 
 EoR 21-cm power spectrum. While all the different frequencies have been collapsed for the 
present analysis, it is necessary to consider the multi-frequency angular power spectrum 
$C_{\ell}(\nu_1,\nu_2)$ or equivalently the three dimensional power spectrum $P(k_{\parallel},k_{\perp})$ 
to quantify the 21-cm signal. We plan to generalize the TGE for this context in future work. 
\section{Acknowledgements}
 We would like to thank the anonymous referee for useful suggestions to improve this paper. SC would like to acknowledge the University Grant Commission, India for providing financial support through Senior Research Fellowship. SSA would like to acknowledge C.T.S, I.I.T. Kharagpur for the use of its facilities. SSA would also like to thank the authorities of the IUCAA, Pune, India for providing the Visiting Associateship programme. AG acknowledge the financial support from the European
Research Council under ERC-Starting Grant FIRSTLIGHT-258942 (PI: L. V. E. Koopmans).


\begin{thebibliography}{99}
\bibitem[\protect\citeauthoryear{Ali, Bharadwaj \& Chengalur}{2008}]{ali08}  
  Ali S. S., Bharadwaj S.,\& Chengalur J.~N., 2008, MNRAS, 385, 2166A
\bibitem[Ali et al.(2015)]{zali15} 
Ali, Z.~S., Parsons, A.~R.,Zheng, H., et al.\ 2015, \apj, 809, 61 
\bibitem[\protect\citeauthoryear{Bernardi et al.}{2009}]{bernardi09} 
 Bernardi, G., de Bruyn, A.~G.,  Brentjens, M.~A., et al.\ 2009, \aap, 500, 965
\bibitem[\protect\citeauthoryear{Begum et al.}{2006}]{begum06} 
 Begum, A., Chengalur, J.~N., \& Bhardwaj, S.\ 2006, \mnras, 372, L33
\bibitem[Bhatnagar et al.(2008)]{bhat08} 
Bhatnagar, S., Cornwell, T.~J., Golap, K., \& Uson, J.~M.\ 2008, \aap, 487, 419 
\bibitem[Bhatnagar et al.(2013)]{bhat13} 
Bhatnagar, S., Rau,U., \& Golap, K.\ 2013, \apj, 770, 91
\bibitem[\protect\citeauthoryear{Bowman, Morales \& Hewitt}{2009}]{bowman} 
 Bowman J.~D., Morales M.~F., \& Hewitt  J.~N.\ 2009, \apj, 695, 183
\bibitem[Chapman et al.(2012)]{chapman12} 
Chapman, E., Abdalla, F.~B., Harker, G., et al.\ 2012, \mnras, 423, 2518
\bibitem[\protect\citeauthoryear{Choudhuri et al.}{2014}]{samir14} 
 Choudhuri, S.,  Bharadwaj, S., Ghosh, A., \& Ali, S.~S.,\ 2014, \mnras, 445, 4351
\bibitem[\protect\citeauthoryear{Choudhuri et al.}{2016}]{samir15} 
 Choudhuri, S.,  Roy, N., Bharadwaj, S., Ali, S.~S., Ghosh, A., \& Dutta,
 P.\ 2016, submitted to MNRAS
 \bibitem[Cornwell \& Perley(1992)]{cornwell92} 
Cornwell, T.~J., \& Perley, R.~A.\ 1992, \aap, 261, 353 
\bibitem[Cornwell et al.(2008)]{cornwell08} 
Cornwell, T.~J.,Golap, K., \& Bhatnagar, S.\ 2008, IEEE Journal of Selected Topics in Signal Processing, 2, 647 
\bibitem[Datta et al.(2010)]{datta10} 
Datta, A., Bowman, J.~D., \& Carilli, C.~L.\ 2010, \apj, 724, 526 
\bibitem[\protect\citeauthoryear {Dillon et al.}{2014}]{dillon14} 
 Dillon, J.~S., Liu, A., Williams, C.~L., et al.\ 2014, PRD, 89, 023002 
\bibitem[\protect\citeauthoryear{Dutta et al.}{2007}]{prasun07}
  Dutta, P., Begum, A., Bharadwaj, S., Chengalur, J.~N.\ 2007, \mnras, 384, L34
\bibitem[\protect\citeauthoryear{Ghosh et al.}{2011{\natexlab{a}}}]{ghosh1} 
 Ghosh, A., Bharadwaj, S., Ali, S.~S., \& Chengalur, J.~N.\ 2011{\natexlab{a}}, \mnras, 411, 2426
\bibitem[\protect\citeauthoryear{Ghosh et al.}{2011{\natexlab{b}}}]{ghosh2} 
 Ghosh, A., Bharadwaj, S.,  Ali, S.~S., \& Chengalur, J.~N.\ 2011{\natexlab{b}}, \mnras, 418, 2584
\bibitem[Ghosh et al.(2012)]{ghosh150} 
Ghosh, A., Prasad, J., Bharadwaj, S., Ali, S.~S., \& Chengalur, J.~N.\ 2012, \mnras, 426, 3295 
\bibitem[\protect\citeauthoryear{Iacobelli et al.}{2013}]{iacobelli13}
 Iacobelli, M., Haverkorn, M., Orr{\'u}, E., et al.\ 2013, \aap, 558, A72
\bibitem[Intema et al.(2009)]{intema09} 
Intema, H.~T., van der Tol, S., Cotton, W.~D., et al.\ 2009, \aap, 501, 1185 
\bibitem[Jeli{\'c} et al.(2008)]{jelic08} 
Jeli{\'c}, V., Zaroubi, S., Labropoulos, P., et al.\ 2008, \mnras, 389, 1319 
\bibitem[Kazemi et al.(2011)]{kazemi11} 
Kazemi, S., Yatawatta,S., Zaroubi, S., et al.\ 2011, \mnras, 414, 1656 
\bibitem [\protect\citeauthoryear{Liu \& Tegmark}{2012}]{liu2}
 Liu, A., \& Tegmark, M.\ 2012, \mnras, 419, 3491
\bibitem [\protect\citeauthoryear{Liu et al.}{2014a}]{liu14a}
 Liu, A., Parsons, A.~R., \& Trott, C.~M.\ 2014a, PRD, 90, 023018
\bibitem [\protect\citeauthoryear{Liu et al.}{2014b}]{liu14b}
 Liu, A., Parsons, A.~R., \& Trott, C.~M.\ 2014b, PRD, 90, 023019
\bibitem[Mao(2012)]{mao12} 
Mao, X.-C.\ 2012, \apj, 744, 29 
\bibitem[Neben et al.(2015)]{neben} 
Neben, A.~R., Bradley, R.~F., Hewitt, J.~N., et al.\ 2015, Radio Science, 50, 614 
\bibitem[Paciga et al.(2011)]{paciga11} 
Paciga, G., Chang,T.-C., Gupta, Y., et al.\ 2011, \mnras, 413, 1174 
\bibitem[\protect\citeauthoryear{Paciga et al.}{2013}]{paciga13}
  Paciga, G., Albert, J.~G., Bandura, K., et al.\ 2013, \mnras, 433, 639 
\bibitem[Parsons et al.(2012)]{parsons12} 
Parsons, A.~R., Pober, J.~C., Aguirre, J.~E., et al.\ 2012, \apj, 756, 165 
\bibitem[\protect\citeauthoryear{Parsons et al.}{2014}]{parsons14}
 Parsons, A.~R., Liu, A., Aguirre, J.~E., et al.\ 2014, \apj, 788, 106 
\bibitem[Platania et al.(1998)]{platinia98} 
Platania, P.,Bensadoun, M., Bersanelli, M., et al.\ 1998, \apj, 505, 473 
\bibitem[Pober et al.(2013)]{pober13} 
Pober, J.~C., Parsons, A.~R., Aguirre, J.~E., et al.\ 2013, \apjl, 768, L36 
\bibitem[Pober et al.(2014)]{pober14} 
Pober, J.~C., Liu, A., Dillon, J.~S., et al.\ 2014, \apj, 782, 66 
\bibitem[Pober et al.(2016)]{pober16} 
Pober, J.~C., Hazelton, B.~J.,Beardsley, A.~P., et al.\ 2016, arXiv:1601.06177 
\bibitem[Seljak(1997)]{seljak97} 
Seljak, U.\ 1997, \apj, 482, 6 
\bibitem[Swarup et al.(1991)]{swarup91} 
Swarup, G., Ananthakrishnan, S., Kapahi, V.~K., et al.\ 1991, Current Science, 60, 95 
\bibitem[\protect\citeauthoryear{Thyagarajan et al.}{2013}]{thyag13}
Thyagarajan, N., Udaya Shankar, N., Subrahmanyan, R., et al.\ 2013, \apj, 776, 6 
\bibitem[\protect\citeauthoryear{Trott et al.}{2012}]{trott1}
 Trott, C.~M., Wayth, R.~B., \& Tingay, S.~J.\ 2012, \apj, 757, 101 
\bibitem[van Haarlem et al.(2013)]{van13} 
van Haarlem, M.~P., Wise, M.~W., Gunst, A.~W., et al.\ 2013, \aap, 556, A2 
\bibitem[\protect\citeauthoryear{Vedantham et al.}{2012}]{vedantham12}
 Vedantham, H., Udaya Shankar, N., \& Subrahmanyan, R.\ 2012, \apj, 745, 176
\end{thebibliography}
\end{document}